\documentclass[%
 reprint,
 superscriptaddress,
 amsmath,amssymb,
 aps,
]{revtex4-1}

\usepackage{graphicx}
\usepackage{dcolumn}
\usepackage{bm}
\usepackage{xcolor}

\begin{document}


\title{Classical information entropy of parton distribution functions
and an application in searching gluon saturation}

\author{Rong Wang}
\email{rwang@impcas.ac.cn}
\affiliation{Institute of Modern Physics, Chinese Academy of Sciences, Lanzhou 730000, China}
\affiliation{School of Nuclear Science and Technology, University of Chinese Academy of Sciences, Beijing 100049, China}


\date{\today}

\begin{abstract}
Entropy or information is a fundamental quantity contained in a system
in statistical mechanics and information theory.
In this paper, a definition of classical information entropy
of parton distribution functions is suggested.
The extensive and supper-additive properties of the defined entropy are discussed.
The concavity is also deduced for the defined entropy.
As an example, the classical information entropy of the gluon distribution
of the proton is presented. There are some particular features of the evolution
of the information entropy in the saturating domain,
which could be used in finding the signals of gluon saturation.
\end{abstract}

\maketitle

\section{Introduction}
\label{sec:intro}

The hadrons are the composite particles made of quarks
and gluons, with complex inner structures.
The one-dimensional momentum distributions
of quarks and gluons of a hadron are described with parton distribution functions
(PDFs) in the infinite momentum frame \cite{Bjorken:1968dy,Feynman:1969ej,Bjorken:1969ja,Gribov:1973jg}.
Thanks to the collinear factorization theorem \cite{Collins:1987pm,Collins:1989gx,Sterman:1995fz},
the PDFs are the universal quantities in different scattering
processes involving the hadron at high energies.
PDFs are of the nonperturbative origin in quantum chromodynamics (QCD)
theory \cite{Yang:1954ek,Gross:1973id,Politzer:1973fx}.
The PDFs at the high energy scale is connected with the nonperturbative
dynamics at the low energy scale, which is peculiarly hard to be calculated.
Nevertheless, PDFs can be well extracted from the experimental measurements
of high-energy reactions, such as the analyses of PDFs done
by CT14 \cite{Dulat:2015mca} and NNPDF \cite{Ball:2010de,Ball:2011uy}.
Nowadays, the extracted PDF data sets are an important tool for
the calculations of high-energy processes involving hadrons
and the simulations of high-energy hadron colliders or fixed-target experiments.

Entropy is an important quantity of a system in thermodynamics.
The second law of thermodynamical physics says that
the entropy can not be reduced during the spontaneous evolution of the system.
According to the famous Boltzmann entropy $S=k_{\rm B}{\rm ln}W$,
the entropy describes the disorder or complexity of the system at the microscopic level.
Here $W$ denotes the number of microstates that correspond to
the same macroscopic thermodynamic state.
The most general formula in statistical mechanics is Gibbs entropy,
as $S=-k_{\rm B}\sum p_{i}{\rm ln}(p_i)$.
The Gibbs entropy turns into the Boltzmann entropy
if all the microstates have the same probability.
The entropy decreases to zero for a perfectly sharp distribution.
The defined entropy in statistical mechanics is the only entropy
that is equivalent to the classical thermodynamic entropy.

The information entropy in information theory was introduced by C. Shannon,
and it is defined as a measure of how much ``choice'' or ``surprise''
is involved in the measurement of a random variable \cite{shannon1948}.
It is a quantification of the expected amount of information
conveyed by identifying the outcome of a random variable.
The definition of Shannon entropy is similar in mathematical form
to that of the Gibbs entropy.
Actually, the Gibbs entropy can be seen as simply the amount of Shannon information
needed to identify the microscopic state of the system, given its macroscopic descriptions.
The information entropy is a useful tool.
It provides an important criterion for setting up probability distributions
on the basis of partial knowledge,
which leads to the maximum-entropy estimate of statistical inference.
The prescription to get the equilibrium distributions of statistical mechanics
by maximizing the Gibbs entropy subject to some constraints
resembles the maximum-entropy principle in statistical inference.
E. Jaynes argued that the statistical mechanics can be taken
as a form of statistical inference rather than as a physical theory
of which the additional assumptions are not contained
in the laws of mechanics (such as ergodicity) \cite{Jaynes:1957zza,Jaynes:1957zz}.

Investigating the inner structure of a hadron
in a statistical view is a novel approach.
How to define an entropy of the inner constituents
inside a hadron in terms of PDFs is an interesting question.
The picture of the partons inside a hadron is ``frozen''
during the short time of measurement.
In the deep inelastic scattering (DIS) process for determining the PDFs,
the quarks probed by the high-energy virtual photon just
resemble the free and real particles: no strongly interactions
or collisions between partons, no appearing and no disappearing
during the short detecting time.
Thus, the definition of a classical entropy of the partons
with PDFs is the primary motivation of this work.

The organization of the paper is as follows.
The derivation of a classical information entropy of parton distribution
function is given in Sec. \ref{sec:entropy-derivation}.
A simple application of the defined entropy is provided
in Sec. \ref{sec:saturation}. Comparisons with other entropies
of quark and gluons are discussed in Sec. \ref{sec:comparisons}.
At the end, a summary is given in Sec. \ref{sec:summary}.

\section{A derivation of a classical entropy for parton distribution function}
\label{sec:entropy-derivation}

The hadron PDFs are the parton number density distributions in the $x$-space.
To study the classical information entropy of PDFs,
a proper definition of the entropy should be made
with any given density distribution.
To ensure the extensive property of the classical entropy,
I find, the definition of the classical information entropy can be given as,
\begin{equation}
\begin{split}
&S\equiv -\left[\int f(x){\rm ln}(f(x))dx -N{\rm ln}(N)+\alpha N\right], \\
&N\equiv \int f(x)dx,
\end{split}
\label{eq:entropy-def-guess}
\end{equation}
where $f(x)$ is the given density distribution which describes a system
and $\alpha$ is an arbitrary constant.
With a simple calculation by this definition, one finds that
the entropy of $k$ times copy of a system equals $k$ times
the entropy of the system, which is written as,
\begin{equation}
\begin{split}
S\left[kf(x)\right]=kS\left[f(x)\right].
\end{split}
\label{eq:extensive-property}
\end{equation}
The extensity of the classical entropy is met.
With the definition in Eq. (\ref{eq:entropy-def-guess}),
the supper-additive property of the entropy is given by,
\begin{equation}
\begin{split}
S\left[f(x)+g(x)\right]\geq S\left[f(x)\right] + S\left[g(x)\right].
\end{split}
\label{eq:additive-property}
\end{equation}
The equality between $S\left[f(x)+g(x)\right]$ and
$S\left[f(x)\right] + S\left[g(x)\right]$ happens only if $g(x)=kf(x)$.
The detailed proof of the supper-additive property is given in the appendix.
With the extensive and supper-additive properties,
the concavity can be easily derived:
$S[\sum_i \lambda_i f_i(x)] \geq  \sum_i S[\lambda_i f_i(x)] = \sum_i \lambda_i S[f_i(x)]$.

In information theory, for a discrete random variable with probability $p_i$,
the information entropy is defined as,
\begin{equation}
\begin{split}
S=-\sum_i p_i {\rm ln}(p_i).
\end{split}
\end{equation}
Let us derive a proper definition of the information entropy
for a continuous random variable $x$,
based on the definition for discrete random variable.
For any given density distribution $f(x)$,
one can construct a probability density distribution $\hat{f}(x)$
by doing the normalization, as $\hat{f}(x)=f(x)/N$ and $N=\int f(x)dx$.
Let us discretize the continuous random variable $x$ with tiny bin width $h$.
If $h$ is small enough, one has the information entropy for
the probability density distribution $\hat{f}(x)$ as,
\begin{equation}
\begin{split}
S[\hat{f}(x)] = -\sum_{i} \hat{f}(x_i)h {\rm ln}\left(\hat{f}(x_i)h\right) \\
= -\sum_{i} \hat{f}(x_i) {\rm ln}\left(\hat{f}(x_i)h\right) h \\
= -\int \hat{f}(x) {\rm ln}\left(\hat{f}(x)h\right) dx.
\end{split}
\label{eq:entropy-normalized-density}
\end{equation}
Replacing $\hat{f}(x)$ with $f(x)/N$, one has,
\begin{equation}
\begin{split}
S[\hat{f}(x)] = -\int \hat{f}(x) {\rm ln}\left(\hat{f}(x)h\right) dx \\
= -\int \frac{f(x)}{N}{\rm ln}\left(\frac{f(x)}{N}h\right) dx \\
= -\frac{1}{N} \left[ \int f(x){\rm ln}(f(x))dx - \int f(x) {\rm ln}(N) dx \right. \\
  \left. + \int f(x) {\rm ln}(h) dx \right] \\
= -\frac{1}{N} \left[ \int f(x){\rm ln}(f(x))dx - N{\rm ln}(N) + N{\rm ln}(h)\right].
\end{split}
\end{equation}
With the assumption of extensive property, one gets,
\begin{equation}
\begin{split}
S[f(x)] = N S[\hat{f}(x)] = \\
= -\left[ \int f(x){\rm ln}(f(x))dx - N{\rm ln}(N) + N{\rm ln}(h)\right].
\end{split}
\label{eq:entropy-def}
\end{equation}
One sees that ${\rm ln}(h)$ is actually the $\alpha$ parameter in Eq. (\ref{eq:entropy-def-guess}).
Since the density distribution $f(x)$ can be greater than 1 in some regions of $x$,
the term ``$-\int f(x){\rm ln}(f(x))dx$'' in the definition can be a negative value.
Therefore the term ``$-N{\rm ln}(h)$'' is important to make sure the entropy is positive,
as long as $h$ is small enough. In the derivation of the entropy,
$h$ should be a very small quantity so that the integral equals the summation
(see Eq. (\ref{eq:entropy-normalized-density})).
In practice, $h$ should be much smaller than the resolution in a measurement.

Entropy is an essential tool to quantify the level of disorder of a system,
or the amount of ``missing'' information needed to determine the microstate
of the system given the macrostate.
We know that a hadron is a complex
system of many partons viewed by a probe at high energy.
Therefore the entropy concept can be applied to the hadron,
and it could be a useful quantity in characterizing the hadron structure.

The maximum entropy principle tells us that
the system is at the maximum entropy for the intestable distributions.
The maximum entropy method is successful in the study
of valence quark distributions of proton \cite{Wang:2014lua}.
The valence quark distributions are determined at the input $Q_0^2$
scale where there are only three valence quarks,
and the number of valence quarks at the scale is known and fixed.
The information entropy for the maximum entropy method
in Ref. \cite{Wang:2014lua} is defined as $-\int f(x){\rm ln}(f(x))dx$,
which is different from the definition in this work.
However, the entropy differences from the variations of distributions
are exactly the same for the definitions in this work and in Ref. \cite{Wang:2014lua},
for $N{\rm ln}(N)-N{\rm ln}(h)$ is a constant
if the number of quarks does not change at the fixed $Q_0^2$ scale.
In the determination of quark distributions with maximum entropy method,
the entropy difference matters instead of the absolute value of the entropy.
Therefore the valence quark distributions determined in Ref. \cite{Wang:2014lua}
are still valid and the same with the new entropy definition in this work.

\section{An application in searching gluon saturation}
\label{sec:saturation}

In this paper, an application of the defined information entropy is illustrated
in the search of the gluon saturation in the proton.
The idea of parton saturation is that the occupation numbers of partons
in the light-cone wave function of a hadron do not grow rapidly
and reach a limiting distribution at small $x$ below the saturation momentum $Q_s$,
which was initiated in the study of parton-parton recombination effect
from the inevitable quanta overlapping
\cite{Gribov:1983ivg,Mueller:1985wy,Mueller:1989st,Mueller:1999wm,McLerran:1993ni,McLerran:1993ka,McLerran:1994vd}.
In theory, the gluon saturation is rigorously guaranteed with the
Jalilian-Marian-Iancu-McLerran-Weigert-Leonidov-Kovner equation
\cite{Balitsky:1995ub,Jalilian-Marian:1997qno,Iancu:2000hn,Weigert:2000gi}
and the saturation solution of the equation is called color glass
condensate (CGC) \cite{Iancu:2000hn,Iancu:2001ad,Jalilian-Marian:2005ccm,Gelis:2010nm}.
A much simple evolution equation which also gives
the gluon saturation is the Balitsky-Kovchegov equation \cite{Balitsky:1997mk,Kovchegov:1999yj,Kovchegov:1999ua,Balitsky:2001re}
with the resummation of fan diagrams (two Pomerons merge into one Pomeron)
added to the Balitsky-Fadin-Kuraev-Lipatov evolution \cite{Lipatov:1976zz,Kuraev:1977fs,Balitsky:1978ic}.
The gluon saturation is not only of such an extraordinary behavior of gluons at high density,
but also important to restore the unitarity upper bound in QCD theory \cite{Froissart:1961ux,Martin:1962rt}.

In experiment, it is a crucial problem and an interesting subject to find the clear signals of gluon saturation
in various kinds of reactions in $p$-$p$, $p$-$A$, $e$-$p$ and $e$-$A$ collisions.
Tremendous efforts from many physicists have been made to uncover the obscure signal of saturation.
The traditional observables in high energy collisions, such as the structure functions and
the cross sections of diffractive and inclusive processes, support more or less
the existence of gluon saturation (see a review \cite{Morreale:2021pnn}).
For the past decade, the more promising methods of two-particle correlations are
developed and suggested to identify the gluon saturation.
These two-particle correlations include the azimuthal correlations of dijet, or
dihadron, or lepton-jet productions in the inclusive or diffractive processes
(see a recent article \cite{Tong:2022zwp} and the literatures therein).
In this work we would like to suggest a new way to probe and identify the
gluon saturation by the evolution pattern of the information entropy over the $Q^2$ scale.

\begin{figure}[htp]
\centering
\includegraphics[width=0.4\textwidth]{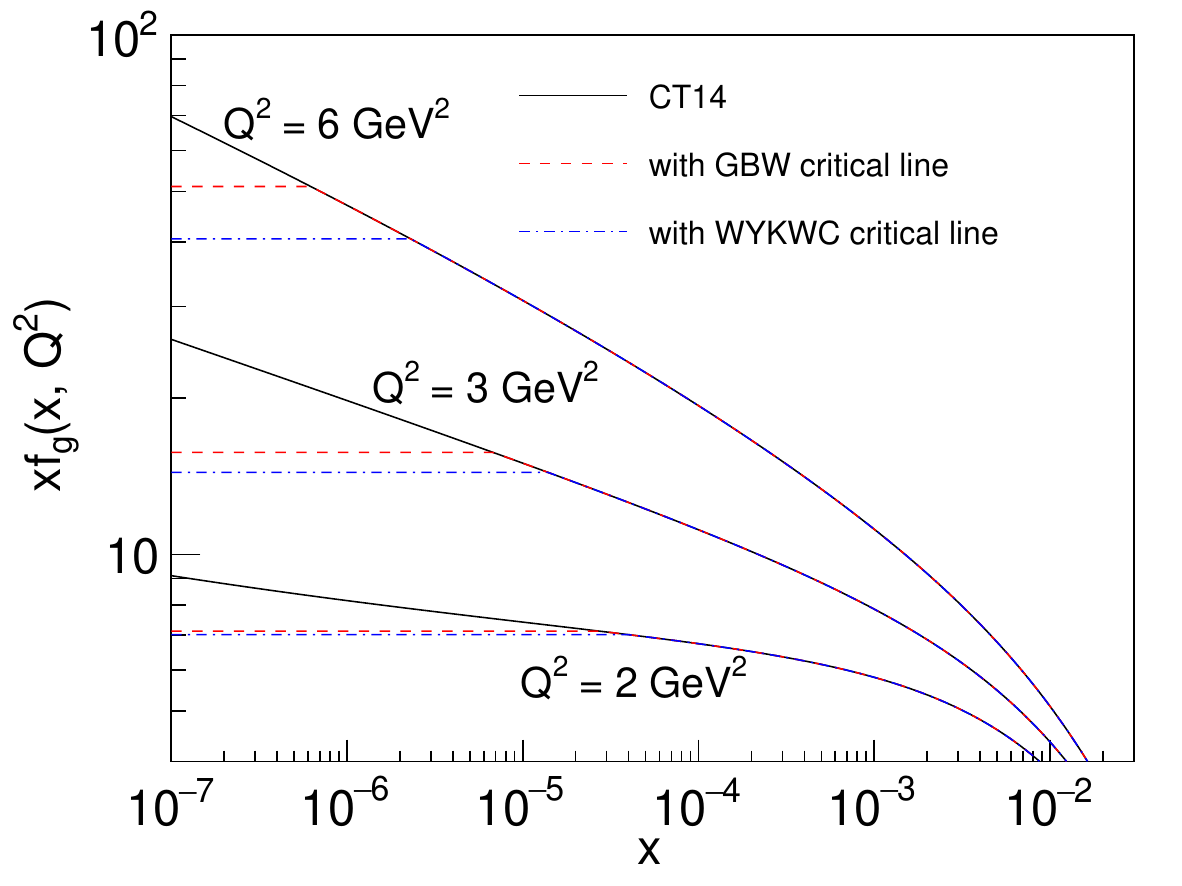}
\caption{
The gluon distributions at small $x$ from CT14 (solid curve) \cite{Dulat:2015mca},
the ideal saturation with GBW critical line (dashed curve) \cite{Golec-Biernat:1998zce,Bartels:2002cj,Kowalski:2003hm},
and the ideal saturation with WYKWC critical line (dash-dotted curve) \cite{Wang:2020stj}.
}
\label{fig:xgluon_saturated}
\end{figure}

\begin{figure}[htp]
\centering
\includegraphics[width=0.4\textwidth]{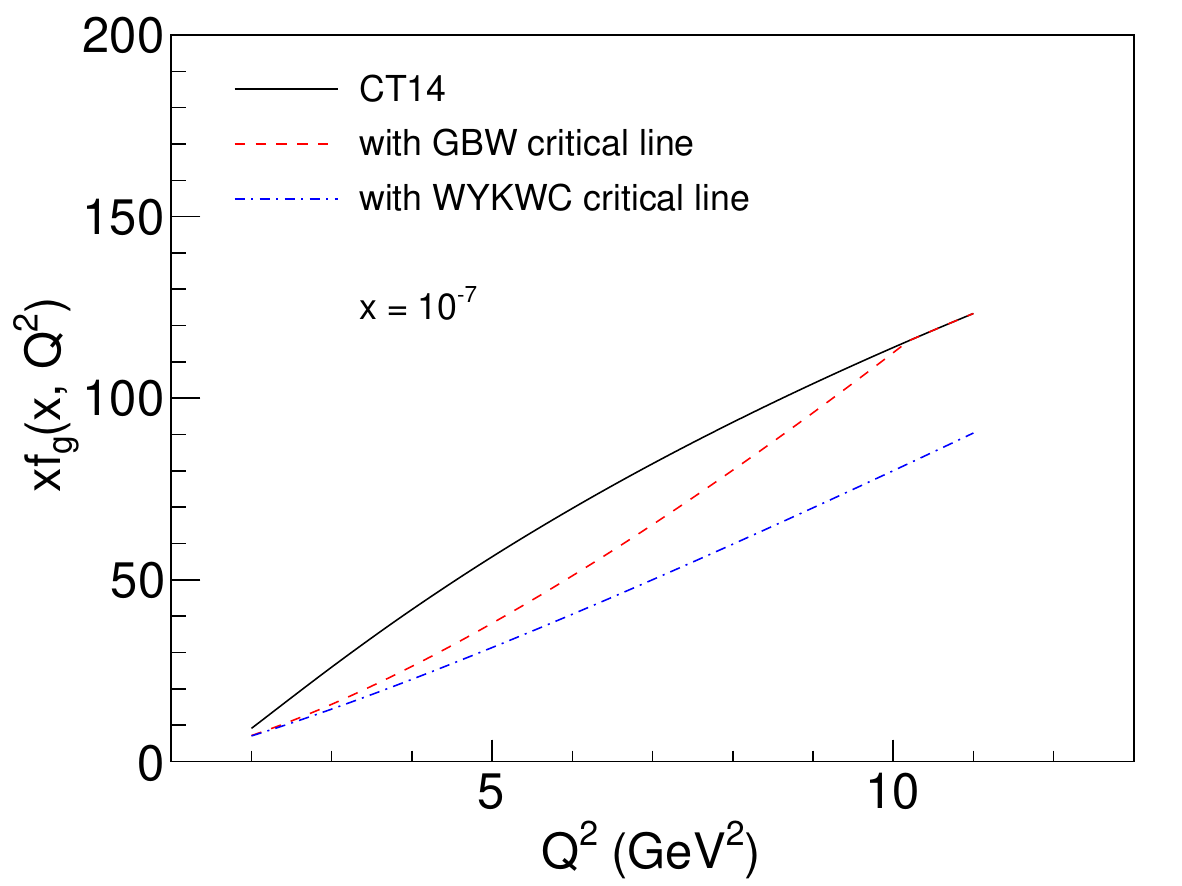}
\caption{
The evolution of gluon distribution at small $x$
in the relatively low $Q^2$ range,
from CT14 PDF (solid curve) \cite{Dulat:2015mca},
the ideal saturation with GBW critical line (dashed curve) \cite{Golec-Biernat:1998zce,Bartels:2002cj,Kowalski:2003hm},
and the ideal saturation with WYKWC critical line (dash-dotted curve) \cite{Wang:2020stj}.
}
\label{fig:xgluon_vs_Q2}
\end{figure}

Before making a prediction for the entropy evolution in the saturation regime,
let us first construct the saturated gluon distributions
at small $x$ below the saturation momentum $Q_s$.
An ideal form of strong gluon saturation is used for the calculations.
According to some QCD models, the gluon density per unit area per unit rapidity
$xg(x,Q^2)$ is a constant in the strong saturation region,
and the gluon density distribution $g(x,Q^2)$ is proportional to $1/x$
\cite{Mueller:1985wy,Mueller:1989st,McLerran:1993ka}.
The saturated gluon distributions at small $x$ are displayed in Fig. \ref{fig:xgluon_saturated},
compared with the normal gluon distribution from CT14 analysis \cite{Dulat:2015mca}.
The sharp and fast transition from nonsaturated distribution
to saturated distribution is assumed across the critical line
between the nonsaturating and saturating regimes.
In this work, the critical lines from GBW's \cite{Golec-Biernat:1998zce,Bartels:2002cj,Kowalski:2003hm}
and WYKWC's \cite{Wang:2020stj} parameterizations are taken.
The WYKWC's critical line is derived from an analytical solution of the simplified BK equation.
The evolutions of the saturated gluon distribution and the CT14 gluon distribution
in the low $Q^2$ range are present in Fig. \ref{fig:xgluon_vs_Q2}.
One sees that below the saturation momentum the saturated gluon distribution increases
approximately linearly with $Q^2$.
Actually this is consistent with the saturation model in the dipole picture \cite{Mueller:1989st,Nikolaev:1990ja}.
The cross section $\sigma^{\rm \gamma^* p}=\sigma_{\rm T} + \sigma_{\rm L}$ is approximately
a constant for small $Q^2$ in the saturation region \cite{Golec-Biernat:1998zce}
(the dipole size $r$ is almost always larger than the saturation scale $R_{0}(x)$ at small $x$).
Eq. (\ref{eq:SF_and_dipole_xsection}) expresses the connection between the $\gamma^{*}p$
cross section and the structure functions. It is clearly shown that the structure functions
or the underlying parton distributions scale linearly with $Q^2$,
in the saturating regime of small $x$ and small $Q^2$.

\begin{equation}
\begin{split}
F_{\rm T,L}(x,Q^2) = \frac{Q^2}{4\pi^2\alpha_{\rm em}} \sigma_{\rm T,L}(x,Q^2),
\end{split}
\label{eq:SF_and_dipole_xsection}
\end{equation}

\begin{figure}[htp]
\centering
\includegraphics[width=0.4\textwidth]{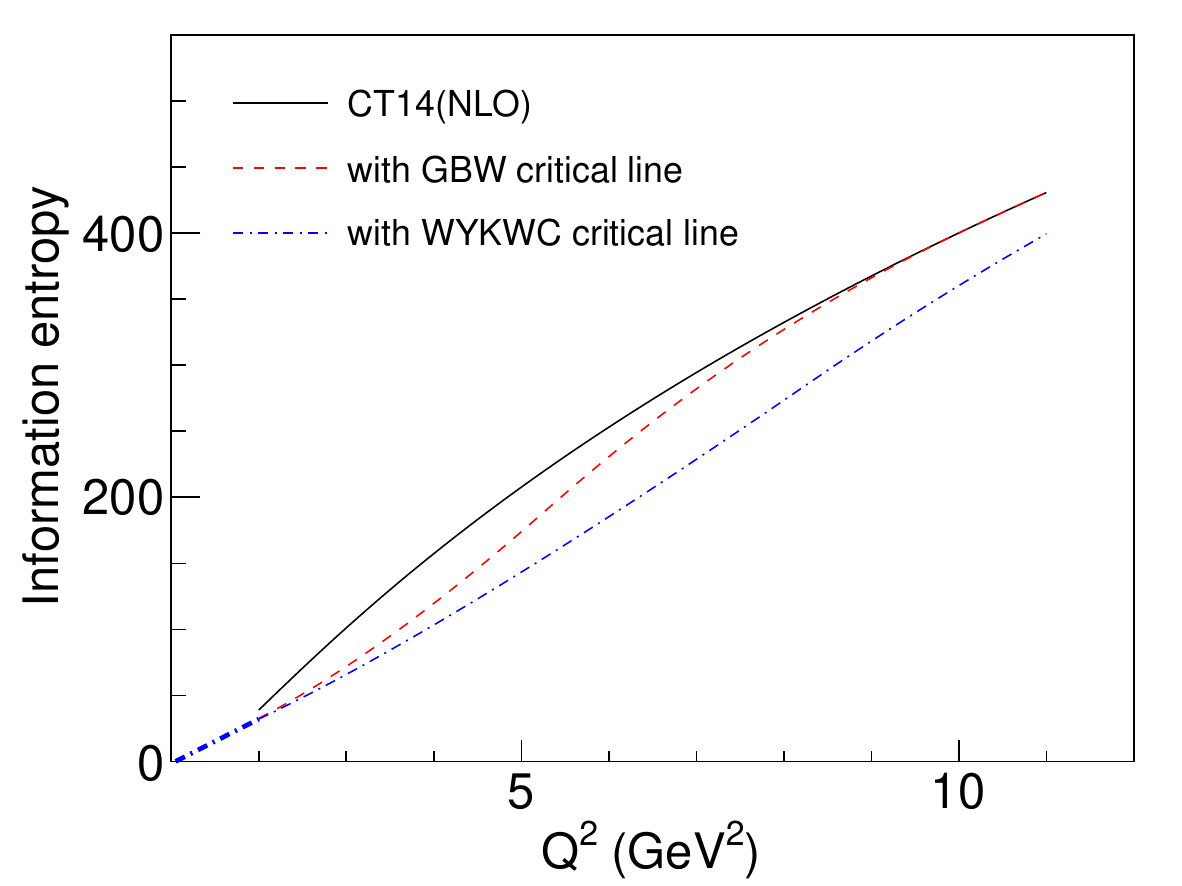}
\caption{
The evolution of information entropy of gluon distribution
in $x$ range [$10^{-7}$, $10^{-6}$],
from CT14 PDF (solid curve) \cite{Dulat:2015mca},
the ideal saturation with GBW critical line (dashed curve) \cite{Golec-Biernat:1998zce,Bartels:2002cj,Kowalski:2003hm},
and the ideal saturation with WYKWC critical line (dash-dotted curve) \cite{Wang:2020stj}.
The thicker dash-dotted line shows the extrapolation of the entropy to low $Q^2$.
}
\label{fig:entropy_evolution_at_small_x}
\end{figure}

\begin{figure}[htp]
\centering
\includegraphics[width=0.4\textwidth]{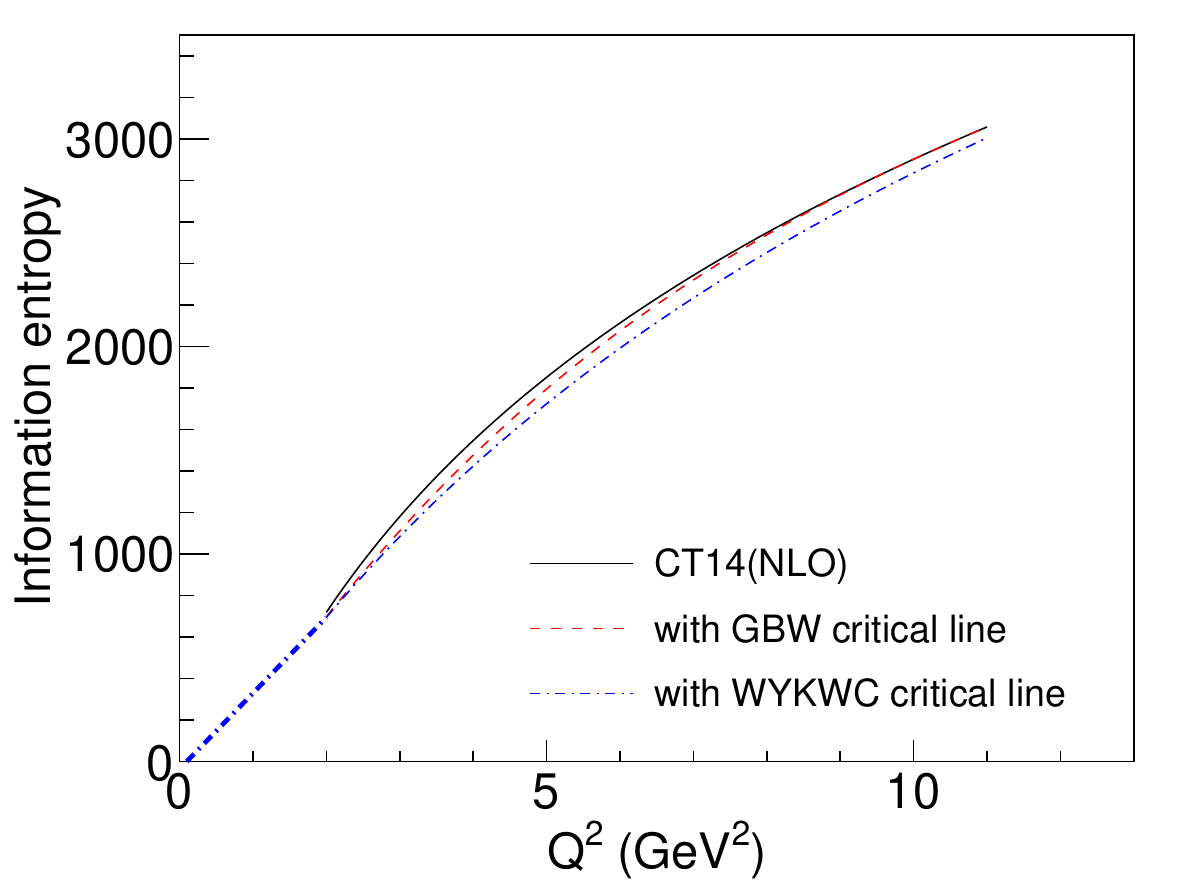}
\caption{
The evolution of information entropy of gluon distribution
in the whole $x$ range [$10^{-7}$, $1$],
from CT14 PDF (solid curve) \cite{Dulat:2015mca},
the ideal saturation with GBW critical line (dashed curve) \cite{Golec-Biernat:1998zce,Bartels:2002cj,Kowalski:2003hm},
and the ideal saturation with WYKWC critical line (dash-dotted curve) \cite{Wang:2020stj}.
The thicker dash-dotted line shows the extrapolation of the entropy to low $Q^2$.
}
\label{fig:entropy_evolution_total_x_range}
\end{figure}

The information entropies are calculated with the saturated and
nonsaturated gluon distributions at different $Q^2$.
The evolutions of the information entropies of gluon distributions with the $Q^2$ scale are
shown in Fig. \ref{fig:entropy_evolution_at_small_x} and \ref{fig:entropy_evolution_total_x_range}.
Fig. \ref{fig:entropy_evolution_at_small_x} presents the information entropy of the gluon distribution
in the limited $x$ range from $10^{-7}$ to $10^{-6}$, while Fig. \ref{fig:entropy_evolution_total_x_range}
presents the information entropy of the gluon distribution in the whole $x$ range from $10^{-7}$ to $1$.
In the calculations, $h$ in Eq. (\ref{eq:entropy-def}) is chosen to be $h=x_{\rm min}=10^{-7}$.
Firstly, one sees that the information entropy of saturated gluon distribution is lower than
that of the nonsaturated gluon distribution (CT14).
Secondly, the information entropy of saturated gluon distribution
in the small-$x$ range increases more or less linearly with the $Q^2$ scale in low $Q^2$ range,
while the entropy of the normal gluon distribution in the small-$x$ range
increases logarithmically with the $Q^2$ scale.
Thirdly, for the information entropy of saturated gluon distribution in the whole $x$ range,
the linear growth of the entropy with $Q^2$ increasing is not obvious,
however the information entropy of saturation is extrapolated to be around zero at $Q^2=0$ GeV$^2$.
These features of the information entropy evolution would be useful
in searching the gluon saturation in experiment.
By evaluating the evolution of entropy in the small-$x$ region,
the overall information from gluon distributions at different $x$
and $Q^2$ are taken into account for identifying the saturation signals.
In addition to the weak $x$-dependence of the gluon distribution,
the linear increase of the information entropy with the $Q^2$ scale is a
quite clear way to distinguish the saturation state of gluons.

While the gluon distribution presents the details of the nucleon structure,
the information entropy provides an overall information in a range of $x$ variable.
Therefore, the evaluated entropy presents smaller statistical uncertainty
(also smaller uncorrelated systematic uncertainty)
compared to the measurement of gluon distribution,
by combining together the experimental data in a $x$-range.
Thus, evaluating the quantity of entropy gives a larger
significance of the saturation signal.
This could be an advantage of applying the defined classical information entropy
to infer the saturation.

\section{Comparisons with entanglement entropy and semiclassical Wehrl entropy}
\label{sec:comparisons}

There are some great progresses on the entanglement entropy of partons recently.
The entanglement entropy at small $x$ is suggested to be
$S(x)={\rm ln}[xg(x)]$, and the DIS probes a maximally entangled state \cite{Kharzeev:2017qzs,Tu:2019ouv}.
The entanglement entropy is also studied with the CGC-Black Hole correspondence \cite{Kou:2022dkw}.
The time evolution of the produced entanglement entropy can be
described with Lipatov's spin chain model \cite{Zhang:2021hra}.
Within this model, the gluon structure function should grow as $xg(x)\sim 1/x^{1/3}$.
The defined classical information entropy in this work
is different from the quantum entanglement entropy of the partons
at a given $x$ and $Q^2$ discussed in Refs. \cite{Kharzeev:2017qzs,Tu:2019ouv}.
In this paper, the Bjorken $x$ is not fixed and treated as a random variable.
The classical entropy in this work quantifies specifically
the ``choice'' of the random variable $x$ in the measurements,
while the quantum entanglement entropy
quantifies the ``choice'' of the parton density or the hadron multiplicity at a fixed $x$.
Moreover, the quantum entanglement entropy is the quantum information entropy,
which quantifies the degree of mixing of the mixed state of a given finite system,
or the ``departure'' of the subsystem from a pure state.
Different entropies characterize the different complexities of the system
in different aspects. Therefore there is no contradiction between
the classical information entropy in this work
and the recently proposed entanglement entropy.

Under the maximally entangled assumption,
the quantum entanglement entropy is related to the gluon distribution,
and it is argued that the entanglement entropy can be accessed
with the Boltzmann entropy of final-sate hadron multiplicity \cite{Kharzeev:2017qzs,Tu:2019ouv}.
In principle, the entanglement entropy is also quite useful
in finding the gluon saturation.
However the correspondence between the initial parton multiplicity
and the final hadron multiplicity depends on the assumption
that no entropy increase during the parton fragmentation process.
Moreover, the hadron multiplicity in the current fragmentation region is
found to be connected to the sea quark distributions
instead of the gluon distribution \cite{Kharzeev:2021yyf}.
Measuring the hadron multiplicity in the target fragmentation region
at small $x$ could provide some valuable information on gluon distribution,
but this kind of measurement is very challenging in experiment.
Because measuring and distinguishing the multiple hadrons in the large rapidity
region (close to the proton beam) require high-performance detectors.
To sum up, (1) theoretically, accessing the gluon entanglement entropy
via final-state hadron multiplicity needs more validations or corrections;
(2) Experimentally, it is challenging to measure all hadrons
in the target fragmentation region.

There is also the semiclassical Wehrl entropy of parton distributions defined
with the Wigner distribution and Husimi distribution \cite{Hagiwara:2017uaz}.
The classical entropy in this work is different from the semiclassical entropy
mainly in the following two aspects.
(1) The entropy in Ref. \cite{Hagiwara:2017uaz} quantifies the complexity of the multi-parton
distributions in transverse phase space $(b_{\perp},k_{\perp})$ with $x$ fixed,
while this paper focuses on the entropy of the probability distribution of $x$.
These two entropies should be applied for different measurements of different variables.
(2) The entropy in Ref. \cite{Hagiwara:2017uaz} is the semiclassical entropy based on
the QCD Husimi distribution, while the entropy defined
in this work is a classical information entropy
based on probability density distribution.

\section{Summary}
\label{sec:summary}

In summary, many classical and quantum entropies of the hadronic system are defined,
and they are the basic and new ways to characterize the complicated hadron structure.
Different entropies quantify the complexities of the hadron structure from
different angles, such as the hadron multiplicity of parton ``liberation'',
the transverse momentum distribution, and the longitudinal momentum distribution.
Therefore different entropies are applied to different physical questions.
For the newly defined information entropy of PDFs, it could be applied in distinguishing
the gluon saturation in experiment.
Although determining PDFs from experimental measurements is a
quite traditional way of probing the nucleon structure, simply
obtaining the gluon distribution is still quite helpful in tackling the saturation
phenomenon at small $x$. Therefore the conventional processes which are sensitive
to the gluon distribution should also be considered for the exploration of gluon saturation,
such as direct photon production \cite{Kumar:2003ue,Campbell:2018wfu,Boettcher:2019kxa},
heavy quarkonium production \cite{ALICE:2017leg,Rezaeian:2013tka,Toll:2012mb},
open charm hadron production \cite{Kelsey:2021gpk},
and jet production \cite{Stump:2003yu,Pumplin:2009nk}
at current hadron colliders and the future
Electron-Ion Collider \cite{AbdulKhalek:2021gbh,Accardi:2012qut}.
With the concept of entropy,
especially the linear evolution of the entropy over the $Q^2$ scale,
the signal of gluon saturation can be clearly identified.
In order to probe the gluon saturation at relatively higher $Q^2$ and larger $x$,
the heavy nuclear target should be used,
as the strong nuclear shadowing and gluon fusion enlarge the domain of gluon saturation.

\begin{acknowledgments}
This work is supported by the National Natural Science Foundation of China
under the Grant NO. 12005266 and
the Strategic Priority Research Program of the Chinese Academy of Sciences
under the Grant NO. XDB34030300.
\end{acknowledgments}

\section*{appendix:}

A simple proof of the super-additive property of the classical information
entropy in Eq. (\ref{eq:entropy-def-guess}) is provided here.
$N_f$, $N_g$ and $k$ are taken to denote respectively the definite
integrals of the density distributions and the ratio between them, as,
\begin{equation}
\begin{split}
N_f\equiv \int f(x) dx, \\
N_g\equiv \int g(x) dx, \\
k\equiv N_g/N_f.
\end{split}
\label{eq:N-k-def}
\end{equation}
Rewriting the formula of the supper-additive property,
we simply need to prove the inequality:
$S[f(x)+g(x)]-S[f(x)]-S[g(x)]\geq 0$.
According to the definition, one has,
\begin{equation}
\begin{split}
S[f(x)+g(x)] - S[f(x)] - S[g(x)] = \\
-\int (f(x)+g(x)) {\rm ln}(f(x)+g(x)) dx \\
+ (N_f+N_g){\rm ln}(N_f+N_g) \\
+\int f(x){\rm ln}(f(x))dx - N_f {\rm ln}(N_f) \\
+ \int g(x) {\rm ln}(g(x))dx - N_g {\rm ln}(N_g) \\
= N_f {\rm ln}\left(\frac{N_f+N_g}{N_f}\right)
+ \int f(x) {\rm ln}\left(\frac{f(x)}{f(x)+g(x)}\right) dx\\
+ N_g {\rm ln}\left(\frac{N_f+N_g}{N_g}\right)
+ \int g(x) {\rm ln}\left(\frac{g(x)}{f(x)+g(x)}\right) dx\\
= \int f(x) {\rm ln}\left(\frac{N_f+N_g}{N_f}\frac{f(x)}{f(x)+g(x)}\right) dx\\
+ \int g(x) {\rm ln}\left(\frac{N_f+N_g}{N_g}\frac{g(x)}{f(x)+g(x)}\right) dx.\\
\end{split}
\label{eq:entropy-increase}
\end{equation}

For any given $g(x)$, one can construct a function $\delta(x)$ as,
\begin{equation}
\begin{split}
\delta(x)\equiv g(x)-kf(x),\\
g(x)\equiv kf(x)+\delta(x),
\end{split}
\end{equation}
in which the $k$ is defined in Eq. (\ref{eq:N-k-def}).
From a simple calculation, one finds that $\int \delta(x) dx =0$.
The function $\delta(x)$ can be viewed as an oscillating function of $x$
which describes the variations of $g(x)$ from $kf(x)$.
With the constructed $\delta(x)$ function and the definition of $k$,
one has,
\begin{equation}
\begin{split}
\int f(x) {\rm ln}\left(\frac{N_f+N_g}{N_f}\frac{f(x)}{f(x)+g(x)}\right) dx \\
= \int f(x) {\rm ln}\left(\frac{(1+k)N_f}{N_f}\frac{f(x)}{(1+k)f(x)+\delta(x)}\right) dx.
\end{split}
\label{eq:one-term-of-entropy-increase}
\end{equation}
The Taylor expansion of $f(x)/[(1+k)f(x)+\delta(x)]$
in terms of $\delta(x)$ is written as,
\begin{equation}
\begin{split}
\frac{f(x)}{(1+k)f(x)+\delta(x)}
= \frac{f(x)}{(1+k)f(x)}
- \frac{f(x)\delta(x)}{[(1+k)f(x)]^2}+\cdot\cdot\cdot \\
=\frac{1}{(1+k)} -\frac{\delta(x)}{(1+k)^2f(x)}+\cdot\cdot\cdot
\end{split}
\end{equation}
By taking the two leading terms of the Taylor expansion,
Eq. (\ref{eq:one-term-of-entropy-increase}) is simplified as,
\begin{equation}
\begin{split}
\int f(x) {\rm ln}\left(\frac{N_f+N_g}{N_f}\frac{f(x)}{f(x)+g(x)}\right) dx \\
= \int f(x) {\rm ln}\left(\frac{(1+k)N_f}{N_f}\frac{f(x)}{(1+k)f(x)+\delta(x)}\right) dx \\
= \int f(x) {\rm ln}\left(1 - \frac{\delta(x)}{(1+k)f(x)}\right) dx.
\end{split}
\end{equation}
Let us look at the Taylor expansion of ${\rm ln}\left(1 - \frac{\delta(x)}{(1+k)f(x)}\right)$,
which is written as,
\begin{equation}
\begin{split}
{\rm ln}\left(1 - \frac{\delta(x)}{(1+k)f(x)}\right) = \\
{\rm ln}(1) - \frac{\delta(x)}{(1+k)f(x)} + \frac{1}{2}\frac{\delta^2(x)}{(1+k)^2f^2(x)} +\cdot\cdot\cdot \\
= - \frac{\delta(x)}{(1+k)f(x)} + \frac{1}{2}\frac{\delta^2(x)}{(1+k)^2f^2(x)} +\cdot\cdot\cdot
\end{split}
\end{equation}
By taking the leading terms of the expansion, one has,
\begin{equation}
\begin{split}
\int f(x) {\rm ln}\left(1 - \frac{\delta(x)}{(1+k)f(x)}\right) dx= \\
\int f(x)\left[- \frac{\delta(x)}{(1+k)f(x)} + \frac{1}{2}\frac{\delta^2(x)}{(1+k)^2f^2(x)}\right] dx \\
= -\frac{1}{1+k} \int \delta(x)dx +\frac{1}{2(1+k)^2} \int \frac{\delta^2(x)}{f(x)}dx.
\end{split}
\end{equation}
Since $\int \delta(x)dx=0$ is provided by definition and
$\delta^2(x)/f(x)$ is always non-negative,
one has,
\begin{equation}
\begin{split}
\int f(x) {\rm ln}\left(\frac{N_f+N_g}{N_f}\frac{f(x)}{f(x)+g(x)}\right) dx \\
= \int f(x) {\rm ln}\left(1 - \frac{\delta(x)}{(1+k)f(x)}\right) dx \\
= -\frac{1}{1+k} \int \delta(x)dx +\frac{1}{2(1+k)^2} \int \frac{\delta^2(x)}{f(x)}dx \\
 \geq 0.
\end{split}
\end{equation}
Similarly, one also has,
\begin{equation}
\begin{split}
\int g(x) {\rm ln}\left(\frac{N_f+N_g}{N_g}\frac{g(x)}{f(x)+g(x)}\right) dx \\
= \int g(x) {\rm ln}\left(\frac{N_f+N_g}{N_g}\left(1 - \frac{f(x)}{f(x)+g(x)}\right) \right) dx\\
= \int g(x) {\rm ln}\left(1 + \frac{\delta(x)}{k(1+k)f(x)}\right) dx.
\end{split}
\end{equation}
By taking the leading term of the Taylor expansion of ${\rm ln}\left(1 + \frac{\delta(x)}{k(1+k)f(x)}\right)$,
one has,
\begin{equation}
\begin{split}
\int g(x) {\rm ln}\left(\frac{N_f+N_g}{N_g}\frac{g(x)}{f(x)+g(x)}\right) dx \\
= \int g(x) {\rm ln}\left(1 + \frac{\delta(x)}{k(1+k)f(x)}\right) dx \\
= \frac{1}{k(1+k)} \int \frac{g(x)\delta(x)}{f(x)} dx \\
= \frac{1}{k(1+k)} \int \left( k \delta(x) + \frac{\delta^2(x)}{f(x)}\right) dx \\
= \frac{1}{k(1+k)} \int \frac{\delta^2(x)}{f(x)} dx \geq 0.
\end{split}
\end{equation}

Now I have proved that,
\begin{equation}
\begin{split}
\int f(x) {\rm ln}\left(\frac{N_f+N_g}{N_f}\frac{f(x)}{f(x)+g(x)}\right) dx \geq 0,
\end{split}
\end{equation}
and
\begin{equation}
\begin{split}
\int g(x) {\rm ln}\left(\frac{N_f+N_g}{N_g}\frac{g(x)}{f(x)+g(x)}\right) dx \geq 0.
\end{split}
\end{equation}
Therefore, the inequality is finally proved, as,
\begin{equation}
\begin{split}
S[f(x)+g(x)] - S[f(x)] - S[g(x)] = \\
= \int f(x) {\rm ln}\left(\frac{N_f+N_g}{N_f}\frac{f(x)}{f(x)+g(x)}\right) dx\\
+ \int g(x) {\rm ln}\left(\frac{N_f+N_g}{N_g}\frac{g(x)}{f(x)+g(x)}\right) dx\\
 \geq 0.
\end{split}
\end{equation}
Note that in the deduction, the function $\delta(x)$ is required to be
a small variation, i.e., $\delta(x)$ is required to be much lower than $f(x)$.
Based on the definition in Eq. (\ref{eq:N-k-def}),
the requirement is met as long as both $g(x)$ and $k$ are small.
In principle, $g(x)$ can be divided into a lot of functions which
are close to zero, as $g(x)=\sum_{i=1}^{i=n} g(x)/n$ with $(g(x)/n)/f(x)<\epsilon$.
Therefore, the supper-additive property of the classical information
entropy is proved, as,
\begin{equation}
\begin{split}
S[f(x)+g(x)] = S\left[f(x)+\sum_{i=1}^{i=n} g(x)/n\right] \\
\geq S\left[f(x)+\sum_{i=1}^{i=n-1} g(x)/n\right] + S[g(x)/n] \\
\geq S\left[f(x)+\sum_{i=1}^{i=n-2} g(x)/n\right] + 2 S[g(x)/n] \\
\cdot\cdot\cdot \\
\geq S[f(x)] + n S[g(x)/n] = S[f(x)] + S[g(x)].
\end{split}
\end{equation}

\bibliographystyle{apsrev4-1}
\bibliography{refs}

\end{document}